\documentclass[aps,rmp,twocolumn,groupedaddress,amssymb,amsfonts,showpacs,showkeys]{revtex4}

\usepackage{color}
\usepackage{times}
\usepackage[dvips]{graphics,graphicx}

\def\half{\frac{1}{2}}

\def\openone{\leavevmode\hbox{\small1\kern-3.8pt\normalsize1}}

\newtheorem{definition}{Definition}

\newtheorem{proposition}{Proposition}

\newcommand{\bproof}{\begin{proof}}
\newcommand{\eproof}{\end{proof}}
\newcommand{\bprop}{\begin{proposition}}

\newcommand{\bdef}{\begin{definition}}

\begin{document}

\title{Entanglement and Subsystems, Entanglement beyond Subsystems, 
\\
and All That}\thanks{Invited contribution to the Proceedings of the
{\em Boston Colloquium for Philosophy of Science on ``Foundations of
Quantum Information and Entanglement''}, Boston, March 23--24, 2006.}

\author{Lorenza Viola}
\email[Electronic address: ]{lorenza.viola@dartmouth.edu}
\affiliation{{Department of Physics and Astronomy, Dartmouth College,\\
6127 Wilder Laboratory, Hanover, NH 03755, USA}}
\author{Howard Barnum}
\email[Electronic address: ]{barnum@lanl.gov}
\affiliation{\mbox{CCS-3}: Modeling, Algorithms, and Informatics, Mail Stop
B256, \\ Los Alamos National Laboratory, Los Alamos, NM 87545, USA}

\date{\today}

\begin{abstract}

Entanglement plays a pervasive role nowadays throughout quantum
information science, and at the same time provides a bridging notion
between quantum information science and fields as diverse as
condensed-matter theory, quantum gravity, and quantum foundations.  In
recent years, a notion of {\em Generalized Entanglement} (GE) has
emerged [H. Barnum {\em et al}, Phys. Rev. A {\bf 68}, 032308 (2003);
L. Viola {\em et al}, Contemp. Math. {\bf 381}, 117 (2005)], based on
the idea that entanglement may be directly defined through expectation
values of {\em preferred observables} -- without reference to a
preferred subsystem decomposition.  Preferred observables capture the
physically relevant point of view, as defined by dynamical,
operational, or fundamental constraints.  While reducing to the
standard entanglement notion when preferred observables are restricted
to arbitrary local observables acting on individual subsystems, GE
substantially expands subsystem-based entanglement theories, in terms
of both conceptual foundations and range of applicability.
Remarkably, the GE framework allows for non-trivial entanglement to
exist within a single, indecomposable quantum system, demands in
general a distinction between quantum separability and absence of
entanglement, and naturally extends to situations where existing
approaches may not be directly useful -- such as entanglement in
arbitrary convex-cones settings and entanglement for indistinguishable
quantum particles.  In this paper, we revisit the main motivations
leading to GE, and summarize the accomplishments and prospects of the
GE program to date, with an eye toward conceptual developments and
implications.  In particular, we explain how the GE approach both
shares strong points of contact with abstract operational quantum
theories and, ultimately, calls for an {\em observer-dependent
redefinition} of concepts like locality, completeness, and reality in
quantum theory.
\end{abstract}

\pacs{03.67.-a, 03.65.Ud, 03.67.Mn, 05.30.-d}
\keywords{Entanglement, locality, observables, convex cones, Lie
algebras}

\maketitle

\section{Introduction}

The first realization that the validity of the quantum superposition
principle in the Hilbert space describing a composite quantum system
may give rise to fundamentally new {\em correlations} between the
constituent subsystems came in the landmark 1935 paper by Einstein,
Podolsky, and Rosen (EPR) \cite{EPR}, where it was shown how the
measurement statistics of observables in certain quantum states could
not be reproduced by assigning definite wave functions to individual
subsystems.  It was in response to the EPR paper that Schr\"odinger,
in the same year, coined the term {\em entanglement}
(``Verschr\"ankung'') to acknowledge the failure of classical
intuition in describing the relationship between the ``parts'' and the
``whole'' in the quantum world \cite{Schrod}:
\begin{quote}
``Whenever one has a complete expectation catalog -- a maximum total
knowledge -- a $\psi$ function -- for two completely separated
bodies,... then one obviously has it also for the two bodies together.
But the converse is not true.  The best possible knowledge of a total
system does not necessarily include total knowledge of all its parts,
not even when these are fully separated from each other and at the
moment are not influencing each other at all.''
\end{quote}

While Bell's strengthening of the original EPR-paradox setting
\cite{Bell} and the subsequent experimental verification of Bell
inequalities \cite{Aspect} irreversibly changed the perception of
entanglement from a property of counterintuitive ``spookiness'' to
(beyond reasonable doubt) an experimental reality, the concept and
implications of entanglement continue to be associated with a host of
physical, mathematical, and philosophical challenges \cite{Pop}.  In
particular, investigation of entanglement in both its {\em
qualitative} and {\em quantitative} aspects has intensified under the
impetus of quantum information science (QIS).  Building on the
discovery of the teleportation and dense coding protocols
\cite{Bennett,BU}, entanglement has nowadays been identified as the
defining resource for quantum communication \cite{BU}, as well as an
essential ingredient for understanding and unlocking the power of
quantum computation \cite{JL}.  Furthermore, entanglement is gaining a
growing status as a key bridging notion between QIS and different
subfields of Physics -- most notably quantum foundations, quantum
statistical mechanics, quantum gravity, and condensed-matter theory.
In spite of continuous progress, however, the current state of
entanglement theory is still marked by a number of outstanding open
problems, which range from the complete classification of mixed-state
bipartite entanglement, to entanglement in systems with continuous
degrees of freedom, and the classification and quantification of
multipartite entanglement for arbitrary quantum states \cite{open}.

At an even more fundamental level, a number of indications have
recently emerged showing that the very {\em definition} of
entanglement as given thus far may be too restrictive to embrace
relevant physical and information-theoretic settings in their full
generality.  From an operational standpoint, the distinction between
entangled and unentangled states of a composite quantum system largely
stems from having acknowledged a separation between ``local''
capabilities -- hereby regarded as a ``cheap'' resource -- as opposed
to arbitrary ``non-local'' capabilities -- which are not readily
available, hence come with a cost: were no operational restriction in
place, then clearly all pure states of the system would be equivalent.
In the conventional approach to entanglement, local capabilities and
degrees of freedom are further (more or less explicitly) identified
with spatially separated distinguishable subsystems.  While such
identification is both natural and adequate for the majority of QIS
settings, compelling motivations for critically reconsidering the
resulting subsystem-based notion of entanglement arise in situations
where the identification of ``local'' resources may not be {\em a
priori} obvious or it may conflict with additional or different
restrictions. A most prominent example in this sense (and one that has
received extensive attention in the recent literature, see e.g.
\cite{eckert,zanardi,wiseman,Kindermann} for representative
contributions) is offered by many-body systems consisting of
indistinguishable (bosonic or fermionic) quantum particles.  Whenever
the spatial separation between the latter is small enough for quantum
statistics to be important, admissible quantum states and observables
are effectively constrained to lie in a proper (symmetric or
antisymmetric) subspace of the respective tensor products of
observable spaces, making the identification of ``local'' subsystems
and operations by far more delicate and, ultimately, ambiguous than in
the standard case.  So, in general, how can entanglement be understood
in an {\em arbitrary} physical system, subject to {\em arbitrary
constraints} on the possible operations we may perform for describing,
manipulating, and observing its states?

Our proposed answer builds on the idea that {\em entanglement is an
inherently relative concept}, whose essential features may be
captured in general in terms of the relationships between different
{\em observers} -- as specified through expectations of quantum
observables in different, physically relevant sets.  In the simplest
instance, distinguished observables in a preferred set determine the
analog of restricted, ``local'' capabilities, as opposed to
unrestricted, ``global'' capabilities embodied by the full observable
space. {\em Generalized Entanglement} (GE) of a quantum state {\em
relative} to the distinguished set may then be defined without without
reference to a decomposition of the overall system into subsystems
\cite{BarnumPRA03,BarnumPRL04}.  That the role of the observer must be
properly acknowledged in determining the distinction between entangled
and unentangled states has been independently stressed by various
authors in various contexts: in particular, the emergence of
distinguished subsystems and of a preferred tensor product structure
has been related to the set of operationally available interactions
and measurements in \cite{qubit,ZLL}, whereas the presence of maximal
entanglement in a state has been directly defined in terms of maximal
fluctuations of fundamental observables in \cite{Klyachko06}.  In
spite of suggestive points of contact, our approach differs from the
above in (at least) two important ways: physically, the need for a
decomposition into distinguishable subsystems is bypassed altogether;
mathematically, the GE notion rests directly (and solely) on {\em
extremality properties of quantum states in convex sets} which are
associated to different observers.  Therefore, GE is both directly
applicable to arbitrary operator languages which may be used to
specify the relevant quantum system, and suitable for investigations
of general operational theories, where convexity plays a key role
\cite{BarnumCones,BarnumFound}.

A more rigorous and thorough development of the GE framework is
available in \cite{BarnumPRA03,ViolaCM05,BarnumCones}; our main goal
here is to informally revisit the key steps and further illustrate
them through examples which may be especially useful at highlighting
conceptual departures from the standard view.

\section{Entanglement and subsystems: The standard view}

In order to motivate and introduce the concept of GE, we begin by
briefly revisiting the standard setup of entanglement theory.
Throughout this paper, our formal treatment of GE properties of
quantum systems will be confined to {\em finite-dimensional} quantum
systems $S$, with associated state spaces ${\cal H}$, dim$\,({\cal
H})=d <\infty$.  In line with Schr\"odinger's original definition, we
furthermore assume that {\em best possible knowledge is available for
$S$}, thus requiring $S$ to be in a {\em pure} state $|\psi\rangle \in
{\cal H}$.  Let ${\cal B}({\cal H})$ denote the space of linear
operators on ${\cal H}$.  The set of (traceless) quantum observables
of $S$ may be naturally identified with the (real) Lie algebra
$\mathfrak{su}(d)\subset {\cal B}({\cal H})$, with Lie bracket
$[X,Y]=i(XY-YX)$. Arbitrary states of $S$ are represented by positive,
normalized elements in ${\cal B} ({\cal H})$ (density operators), with
$\rho=|\psi\rangle\langle \psi|$ for a pure state.  The set $\Upsilon$
of density operators is a compact convex subset of ${\cal B} ({\cal
H})$: thus equivalently, $\rho$ is pure if it is an {\em extreme}
point of $\Upsilon$, or the purity of $\rho$ is maximal, $P(\rho)$=
Tr($\rho^2)=1$.

The essential intuition on which the GE notion builds may be
appreciated starting from the simplest instance of a composite quantum
system, namely a pair of two-dimensional subsystems (Alice and Bob
henceforth), living in a four-dimensional complex space
\begin{equation}
{\cal H}\equiv {\cal H}_{\tt AB} ={\cal H}_{\tt A} \otimes {\cal
H}_{\tt B} \:,
\label{stp}
\end{equation}
where dim$\,({\cal H}_A)=$ dim$\,({\cal H}_B)=2$.  Any joint pure
state of Alice and Bob which is {\em separable}, that is, able to be
expressed in the form
\begin{equation}
|\Psi\rangle_{\tt AB}= |\psi\rangle_{\tt A}
\otimes |\phi\rangle_{\tt B} \equiv |\psi\phi\rangle_{\tt AB}\,,
\label{prod}
\end{equation}
for suitable subsystem states of Alice and Bob alone, is unentangled.
Let $\{ |0\rangle, |1\rangle \}$ denote an orthonormal basis in
${\mathbb C}^2$ (computational basis), as usual.  Then the following
{\em Bell states} are well known to be (maximally) entangled
\cite{Bell},
\begin{eqnarray}
&& |\Psi^\pm\rangle_{\tt AB} =\frac{|00\rangle_{\tt AB} \pm
|11\rangle_{\tt AB}}{\sqrt{2}} \,, \nonumber \\ 
&& |\Phi^\pm\rangle_{\tt AB}
=\frac{|01\rangle_{\tt AB}  \pm |10\rangle_{\tt AB} }{\sqrt{2}}\,.
\label{Bell}
\end{eqnarray}
What distinguishes, at an operational level, states of the form
(\ref{prod}) from states of the form (\ref{Bell})?  While the answer
to this question may be phrased in different ways in principle, it
ultimately rests on the distinction between what Alice and Bob may
accomplish in terms of purely {\em local} resources as opposed to
arbitrary non-local ones.  Let, in particular, 
\begin{equation}
\Omega_{loc} = \mbox{span}_{\mathbb R} \{ A \otimes \openone, 
\openone \otimes B \,|\, A=A^\dagger, B=B^\dagger \}
\label{locobs}
\end{equation}
denote the set of local traceless observables on ${\cal H}_{\tt AB}$,
so that a generic unitary transformation generated by observables in
$\Omega_{loc}$ is of the form $U_{\tt AB}=U_{\tt A}\otimes U_{\tt B}$.
(The traceless condition excludes the identity operator from the
distinguished subspace of observables; it is a somewhat arbitrary
decision whether or not to do this, as it has no effect on the convex
structure of the set of reduced states induced by normalized states,
since all states take the same value on it.  Considering the reduced
states as a base for a cone, however, is equivalent to reintroducing
the identity operator, and considering the reduced states induced by
the states in the cone of {\em unnormalized} states.)  Note that
$\Omega_{loc}$ may be identified with the Lie algebra
\begin{equation}
\Omega_{loc} \simeq \mathfrak{su}(2)_{\tt A} \oplus 
\mathfrak{su}(2)_{\tt B} =\mbox{span}_{\mathbb R} \{\sigma_a \otimes 
\openone, \openone \otimes \sigma_b\}\,,
\label{locobsp}
\end{equation}
where $\sigma_a, a \in \{x,y,z\}$, denote spin Pauli operators on
${\mathbb C}^2$.  Then no Bell state may be reached starting from a
product state as in (\ref{prod}) solely by application of operators
generated by local observables.  Alternatively, imagine that the state
of Alice and Bob is to be determined based only on access to
expectation values of observables \cite{note1}.  Then any pure product
state is completely specified by knowledge of the expectation values
$\langle \sigma_a \rangle_{\tt A,B}$ on each subsystem, whereas
knowledge of the same expectations {\em cannot} distinguish a Bell
state from a {\em mixture} of pure states - containing no
entanglement.  For instance,
\begin{equation}
|\Phi^-\rangle_{\tt AB} \equiv_{loc} \frac{|01\rangle_{\tt AB}\langle
01| + |10\rangle_{\tt AB}\langle 10| }{2}\,,
\end{equation}
where equivalence means indistinguishability by access to expectations
of restricted (local) observables.  In order to distinguish, knowledge
of appropriate {\em correlations} is required, that is expectations of
non-local observables like $\sigma_x \otimes \sigma_x$ in the above
example.  An equivalent characterization of entanglement may be
obtained in terms of purity of Alice and Bob subsystem states, as
given by the corresponding reduced density operators: pure product
states are precisely those states for which {\em each} subsystem
remains pure.  To state it differently, pure entangled states are
those pure states whose {\em reduced} states -- i.e. the expectation
values they determine for all {\em one-party} observables only --are
{\em non-extremal}, i.e. mixed.

Even in the simplest setting under consideration, it is essential to
acknowledge that the characterization of a pure state in ${\cal H}$ as
entangled or not is unambiguously defined only after a {\em fixed}
tensor decomposition has been chosen among the distinct ones that
${\cal H}$ can {\em a priori} support (as long as its dimension $d$ is
a non-prime integer).  By its very nature, (standard) entanglement is
{\em relative} to a preferred subsystem decomposition -- capturing the
specific way in which $S$ is viewed as constituted of its parts.
Physically, one may expect that what makes a given factorization
preferred among others be naturally linked to the set of operations
which are deemed as practically available for control and observation.
Indeed, in the bipartite setting discussed above, availability of
arbitrary observables in $\Omega_{loc}$ may be intuitively (and
formally) related to the identification of local degrees of freedom
associated with subsystems ${\tt A}$ and ${\tt B}$ -- as described in
general by appropriate mutually commuting associative algebras of
operators \cite{qubit,ZLL}.  Suppose, however, that for whatever
reason the rule distinguishing what is ``cheap'' from what is not
changes -- in particular, suppose that accessible observables on
${\cal H}={\mathbb C}^4$ are arbitrary linear combinations in the
following set:
\begin{equation}
\Omega'_{loc} = \mbox{span}_{\mathbb R} \{ \sigma_x \otimes \sigma_x,
\sigma_z \otimes \sigma_z, \sigma_x \otimes \sigma_y, \sigma_y \otimes
\sigma_z \} \,.
\label{locobs2}
\end{equation}
Then expectation values of observables in $\Omega'_{loc}$ are clearly
sufficient to completely specify Bell states as given in
Eq.~(\ref{Bell}), whereas pure states of the form
$|\psi\phi\rangle_{\tt AB}$ may become ``locally'' indistinguishable
from mixtures, for instance
\begin{equation}
|00 \rangle_{\tt AB} \equiv_{loc'} \frac{|00\rangle_{\tt AB}\langle
00| + |11\rangle_{\tt AB}\langle 11| }{2}\,,
\end{equation}
where equivalence has the same meaning as before under
$\Omega'_{loc}$.  Accordingly, Bell states should now be regarded as
{\em un}-entangled with respect to the relevant set of local
capabilities.  In fact, it is possible to show in this case
\cite{qubit,ZLL} that a well-defined decomposition of ${\cal H}$ into
subsystems still exists relative to the new local set $\Omega'_{loc}$,
\begin{equation}
{\cal H} ={\cal H}_{\tt \chi} \otimes {\cal H}_{\tt \lambda} \:,
\label{virtual}
\end{equation}
where $\chi$, $\lambda$ specify ``virtual'' subsystems corresponding
to eigenvalue $\chi \in \{\Psi,\Phi\}$, and $\lambda \in \{+,-\}$,
respectively -- so that, for instance, $|\Psi^+\rangle \simeq
|\Psi\rangle_{\tt \chi} \otimes |+\rangle_{\lambda}$, and so on.

While the above examples nicely demonstrate how, even within the
standard ``subsystem-based'' framework, the notion of entanglement is
strongly observer-dependent, a closer scrutiny rapidly leads to the
following deeper questions:
\begin{quote}
{Are subsystems general and flexible enough to capture the relativity
of entanglement in full?  Once a preferred observer is specified
through the identification of a distinguished observable set, is it
always possible to relate such an observer to the emergence of
preferred subsystems?  And, whether it may be possible or not, is it
always necessary or useful?}
\end{quote}

Several concrete examples may be adduced toward showing the
inadequacy of subsystem-based entanglement in settings involving
operational or fundamental constraints more general than the ones
implied by the standard framework
\cite{BarnumPRA03,SommaPRA04,ViolaCM05,Ortiz,SommaIJP}.  Within the
four-dimensional state space considered so far, imagine for instance
that available operations are subject to a {\em conservation
law}, say conservation of the total (pseudo)spin angular momentum
along a given axis, $S_z=(s_z\otimes \openone + \openone\otimes
s_z)/2$, with $s_a=\sigma_a/2$. Assume we describe the corresponding
observer in terms of the (smallest) Lie algebra of observables
commuting with $S_z$,
\begin{eqnarray}
\Omega^z_{loc} = \mbox{span}_{\mathbb R} \{ &\hspace*{-2.3mm} s_z
\hspace*{-2.3mm}& \otimes \openone, \openone \otimes s_z, \sqrt{2}
(s_x \otimes s_x +s_y \otimes s_y), \nonumber \\
&\hspace*{-2.3mm}\sqrt{2} \hspace*{-2.3mm}& \,(s_x \otimes s_y -s_y
\otimes s_x) \} \simeq \mathfrak{u}(2) \,,
\label{u2}
\end{eqnarray}
where an orthonormal Hermitian basis has been used for later
reference.  On one hand, because only a subset of local observables
are included, only a proper subspace of product states in ${\cal
H}_{\tt AB}$ may expected to remain ``unentangled'' given such
capabilities.  On the other hand, although in analogy with
$\Omega'_{loc}$ expectations of observables in $\Omega^z_{loc}$
continue to determine states like $|\Psi^\pm\rangle$, it is also clear
that not all four Bell states are now on the same footing, as long as
$s_z \otimes s_z$ is not included in the distinguished set.  Hence,
$\Omega^z_{loc}$ corresponds neither to the factorization in
Eq.~(\ref{stp}) nor to that in Eq.~(\ref{virtual}). Yet, one would
still like a natural and meaningful notion of entanglement to exist,
based on the identification of ``local'' resources as the ones
described by $\Omega^z_{loc}$.

Aside from being desirable on fundamental grounds, compelling
motivations for consistently describing a distinguished
$\mathfrak{u}(2)$ observable algebra (and higher-dimensional
generalizations) arise in the context of defining entanglement in
systems of {\em indistinguishable fermions} -- in which case the
latter is reinterpreted in a second-quantized language and the
$S_z$-constraint is imposed by fermion-number conservation (see
Sect. IVD). In general, quantum indistinguishability constrains the
admissible fermionic states to the fully anti-symmetric subspace of
${\cal H}$, preventing a direct identification between particles and
subsystems in the standard tensor product sense.  While factorizations
into distinguishable subsystems can still be defined in terms of
appropriate sets of {\em modes}, the choice of preferred modes may be
problematic in situations where different sets (e.g., spatial and
momentum modes) are equally relevant to the description.  Finally, the
possibility that the operator {\em language} which describes the
problem may itself be changed -- for instance via isomorphic mappings
between spin and fermion operators like the Jordan-Wigner
transformation \cite{batista} -- further adds to the intricacy of
applying the standard entanglement framework to general quantum
many-body systems.

Can entanglement be {\em directly defined in terms of distinguished
physical observables}, irrespective of and without reference to a
preferred subsystem decomposition?

\section{Entanglement beyond subsystems: The concept of generalized entanglement}

The generalized entanglement setting we are seeking must satisfy two
essential requirements: it must both (i) reduce to the ordinary
framework in well-defined limiting situations, and (ii) identify the
presence or absence of entanglement independently of the specific
operator language used to describe the system $S$.  The key intuition
is to generalize the characterization of pure-state entanglement as
{\em relative mixedness under restricted capabilities}.

While we refer to \cite{BarnumCones} for a more detailed and
mathematically more rigorous account, all relevant GE settings are
subsumed as special cases of a general definition based on {\em
distinguished cone-pairs}.  Recall that a (finite-dimensional) convex
cone is a proper subset of a (finite-dimensional) real vector space
closed under multiplication by nonnegative scalars and under addition.
In our usage, the term will refer to {\em regular} convex cones, that
is, cones that are pointed (contain no subspace other than $\{0\}$),
generating (so that the linear span of the cone is the full ambient
vector space) and topologically closed.  Let $V,V^\ast$ respectively
denote a real linear space and its dual, that is, the space of all
linear functionals from $V$ to ${\mathbb R}$.  Given a convex cone $C
\subseteq V$ and a distinguished functional $\lambda \in V^\ast$, we
associate states to normalized elements in $x \in C$, satisfying
$\lambda(x)=1$.  We require that $\lambda$ separate $C$ from $-C$,
equivalently that the only element $x \in C$ for which $\lambda(x)=0$
is $x=0$, and we also require that $\lambda(C) \ge 0$.  These
conditions are imposed so that (given the regularity of the cone) the
set $\hat{C}$ of normalized states in $C$ is a {\em compact convex
set}.  Also, each element of $C$ can be written as $\alpha x$ for some
unique $x \in \hat{C}$, $\alpha > 0$ (in other words, $\hat{C}$ is a
{\em base} for the cone).  Thus $C$ may be thought of as the set of
{\em unnormalized} states: multiples of the normalized states which
belong to $\hat{C}$.  We will use the term {\em extremal state} to
mean an extremal element of $\hat{C}$, in the standard sense that it
cannot be written as a nontrivial convex combination of two distinct
elements of $\hat{C}$.  But where it is clear that the state is in
general unnormalized, we may use the term extremal state to refer to
an unnormalized state belonging to an extremal ray of $C$, that is, a
nonnegative multiple of an extremal normalized state~\cite{warn}.

The operational interpretation of the above construction also requires
us to consider the dual cone $C^\ast = \{\alpha \in V^* |$ $ \alpha(x)
\ge 0 ~\forall x \in C \}$, which consists of the functionals
nonnegative on $C$.  $C^\ast$ is interpreted as the set of possible
``effects,'' corresponding to ``unnormalized'' measurement outcomes,
for states in $C$; $\alpha (x)$, which is nonnegative for all $x \in
C$, is interpreted as an (unnormalized) probability for outcome
$\alpha$ when the state is $x$.  Observe that the distinguished
functional $\lambda$ (often called the ``unit'' or ``order unit'')
belongs to $C^*$: It is the measurement outcome that has probability
$1$ in all states.  The interval $[0,\lambda] \subseteq C^*$ defined
by $\{\alpha \in C^* \,|\, \alpha \in C, \lambda - \alpha \in C\}$
corresponds to outcomes normalized so that they can appear in a
measurement, as it is easily verified that $[0,\lambda]$ is precisely
the set of functionals such that for all $x \in \hat{C}$, $0 \le
\alpha(x) \le 1$, enforcing that probabilities for measurement
outcomes lie in $[0,1]$.  Measurements with a finite set of outcomes
correspond to finite {\em resolutions of the unit} into elements of
$[0,\lambda]$: finite sequences $\alpha_i$ of elements of
$[0,\lambda]$ such that $\sum_i \alpha_i = \lambda$, which, as is
easily checked, enforces that the probabilities of the measurement
outcomes $i$ sum to $1$ in all states, i.e.  $\sum_i \alpha_i(x) = 1$.
Note that for many (but not all) purposes it may be easier to ignore
this additional derived structure of a dual cone and a unit interval
within it, viewing $V^*$ as simply a set of real-valued
linear-in-the-states ``observables'' and the states in $C$ as
determining their expectation values.

GE may be defined once a pair of distinguished state-sets --
generalizing the distinction between states as accessible to a
``local'' as opposed to a ``global'' observer -- and a choice of an
appropriate normalization-preserving linear map -- generalizing the
notion of computing the reduced density operator for bipartite systems
-- are specified:

\vspace*{1mm}

{\bf Definition}. Let $V,W$ be real linear spaces equipped with
distinguished convex cones $C \subset V, D\subset W$ and positive
semidefinite linear functionals $\lambda\in C^\ast, \tilde{\lambda}
\in D^\ast$ satisfying the requirements discussed above.  
Let $\pi: V \rightarrow W$ be a normalization-preserving
linear map taking $C$ onto $D$ that is, 

i) $\pi(C)=D$, and 

ii) $\pi( \{x\in V\,|\,\lambda (x)=1 \}) = \{ y \in W\,|\, \tilde{\lambda}(y)
=1 \}$.

A pure (extremal) state $x \in C$ is {\em generalized unentangled
relative to D} if its image $\pi(x)$ is pure (extremal) in $D$. A
mixed (non-extremal) state in $z \in C$ is generalized unentangled
relative to $D$ if $z =\sum_a w_a x_a$, for positive numbers $w_a \geq
0$ and extreme points $x_a \in C$ whose images $\pi(x_a)$ are extreme
in $D$.

\vspace*{1mm}

Note that this definition applies in general to unnormalized states,
but includes normalized ones.  In case the states are normalized, it
coincides with a natural restricted definition, in which $C, D$ are
replaced by $\hat{C}, \hat{D}$ above, and extremality is taken in the
convex-sets sense as extremality in $\hat{C},\hat{D}$.  Condition ii)
on the map $\pi$ guarantees that this makes sense, by ensuring that
$\hat{C}$ is mapped to $\hat{D}$.

\subsection{GE settings}

In physical applications, the distinguished functional $\lambda$ is
identified with the trace map, and the reduced state-set associated
with $D$ is obtained by selecting a preferred subspace $\Omega \subseteq
{\cal B}({\cal H})$ of ``local'' observables for $S$.  In particular,
the above general definition may be specialized to the following
relevant entanglement settings \cite{BarnumCones}:

\vspace{1mm}

{\bf GE1: Distinguished quantum observables setting}. $C$ is
isomorphic to the convex cone ${\Upsilon}$ of quantum states on ${\cal
H}$ (positive normalized functionals $\eta$ induced by positive
multiples of density operators), and $V^\ast$ is viewed as the
subspace of Hermitian operators in ${\cal B}({\cal H})$, with
$X[\eta]= \mbox{Tr}(\rho_\eta X)$ giving the expectation value of
observable $X$ in state $\rho_\eta$.  Elements in $V=(V^\ast)^\ast$
are arbitrary linear functionals defined so that $\xi[X]=X[\xi]$, for
all $X\in V^\ast$, $\xi \in V$.  Let the distinguished observable
space be any real linear subspace $\Omega\equiv W^\ast \subset
V^\ast$.  Then the cone $D \subset W$ of $\Omega$-reduced unnormalized
states may be obtained as the image of $C$ under the map $\pi: V
\mapsto W$ which restricts elements of $V$ to observables in the
distinguished set, i.e. for $\eta \in V$, $\pi(\eta)$ is defined by
the condition that for all $X \in W^*=\Omega$, $\pi(\eta)[X] =
\eta(X)$.  Thus, elements in $D$ may be seen as lists of expectation
values for observables in $\Omega$.

\vspace{1mm}

{\bf GE2: Hermitian-closed operator subspace setting}. In this case,
the distinguished observable space $\Omega$ consists of the Hermitian
operators in a linear subspace ${\cal V}\subset {\cal B}({\cal H})$
containing the identity and closed under Hermitian conjugation. This
formulation turns out to be fully equivalent to the GE1-setting
defined above.

\vspace{1mm}

{\bf GE3: Lie-algebraic setting}. In situations where the Hilbert
space ${\cal H}$ of $S$ may be identified with the representation
space of a (Hermitian-closed) Lie algebra $\mathfrak{h}$, it is
natural to identify $\Omega = \mathfrak{h}$, with the corresponding
cone $D$ of reduced $\mathfrak{h}$-states consisting of linear
functionals on $\mathfrak{h}$ induced from positive normalized
operators in ${\cal B}({\cal H})$ upon restriction to $W^\ast=\Omega$.
By construction, this setting is a special case of GE1. Under the
additional assumptions that $\mathfrak{h}$ is semi-simple and acts
irreducibly on ${\cal H}$, this setting leads to an explicit
characterization of the set of pure generalized unentangled states,
which are identical with the set of so-called {\em generalized
coherent states} (GCSs) of $\mathfrak{h}$ \cite{BarnumPRA03,Pere}.

\vspace{1mm}

{\bf GE4: Associative-algebraic setting}. Here, the distinguished set
$\Omega$ consists of the Hermitian elements of an associative
sub-algebra ${\cal A}$ of ${\cal B}({\cal H})$. This case may also be
seen as a special instance of GE1.  Under the additional assumption
that ${\cal A}$ partitions into a collection of independently
accessible, mutually commuting associative sub-algebras $\{ {\cal A}_i
\}$ satisfying $\otimes_i {\cal A}_i \simeq {\cal B}({\cal H})$ (or
generalizations where a proper ``coding'' subspace ${\cal C} \subset
{\cal H}$ is considered), this setting is directly relevant to
identifying observable-induced subsystem-decompositions as in \cite{ZLL}.

\subsection{GE measures}

As with standard entanglement, no single measure can, in general,
uniquely characterize the GE properties of a state.  However, 
the observation
that standard pure-state entanglement translates into mixedness (loss
of purity) of the reduced subsystem states naturally suggests
seeking a way to quantify the degree of purity relative
to the distinguished observable set.

Let $\Omega$ be a Hermitian-closed set, and let $\{ X_\alpha \}$ be 
an orthonormal basis of $\Omega$, $\mbox{Tr}(X_\alpha X_\beta)
=\delta_{\alpha\beta}$. Then for every density operator $\rho$ on
$S$, the projection of $\rho$ onto $\Omega$ may be defined via
$$ {\cal P}_\Omega(\rho) = \sum_\alpha \mbox{Tr}(\rho X_\alpha)
X_\alpha \,.$$ 
\noindent
${\cal P}_\Omega(\rho)$ may be regarded as the ``$\Omega$-reduced''
density operator associated to $\rho$, although (unlike in the
standard multipartite case) ${\cal P}_\Omega(\rho)$ is not necessarily
positive semidefinite: the positivity of $\mbox{Tr}({\cal
P}_\Omega(\rho) X)$ does not in general hold for all positive
semidefinite observables $X$; although if we included the identity in
$\Omega$, it is assured for those $X$ belonging to $\Omega$.

\vspace*{1mm}

{\bf Definition}. For any density operator $\rho \in {\Upsilon}$, the
{\em purity of $\rho$ relative to $\Omega$} ($\Omega$-purity) is given
by the squared length of its $\Omega$-reduced density operator that is,
\begin{equation}
P_\Omega(\rho)=\mbox{Tr}({\cal P}_\Omega(\rho)^2)=
\sum_\alpha |\mbox{Tr}(\rho X_\alpha)|^2 \,.
\label{p1}
\end{equation}
In particular, if $\rho=|\psi\rangle\langle \psi|$ is pure, we simply
write
\begin{equation}
P_\Omega(|\psi\rangle)=\sum_\alpha |\langle \psi |X_\alpha |\psi
\rangle|^2\,.
\label{p2}
\end{equation}

By construction, $0\leq P_\Omega(\rho) \leq P(\rho)\equiv
\mbox{Tr}(\rho^2)$.  Also note that $P_\Omega(|\psi\rangle)=1/d+
P_{\Omega_0}(|\psi\rangle)$, where $\Omega_0 \subseteq \Omega$ denotes
the distinguished traceless sector.  It is
often convenient to discard the constant trace-contribution, in which
case the common normalization constant for the (traceless) $X_\alpha$
is adjusted so to rescale the maximum of $P_{\Omega_0}$ to $1$. In the
Lie-algebraic setting GE3, the $\mathfrak{h}$-purity so constructed is
automatically invariant under arbitrary unitary transformations in the
Lie group generated by $\mathfrak{h}$, as is desirable on physical
grounds.

The usefulness of the $\Omega$-purity as a measure of {\em pure-state}
GE in various settings is summarized in the following Theorem, proved
in \cite{BarnumPRA03} and \cite{BarnumCones}:

\vspace*{1mm}

{\bf Theorem}. i) In the {\em irreducible} Lie-algebraic and
associative-algebraic settings, a pure state $|\psi\rangle\in {\cal
H}$ is generalized unentangled relative to $\Omega$ if and only if
$P_\Omega(|\psi\rangle)$ is maximal.
ii) In the Hermitian-closed operator setting (including {\em
reducibly} represented operator algebras), states with maximal
$P_\Omega$-purity are generalized unentangled relative to $\Omega$.

\vspace*{1mm}

A complete characterization of the relationship between maximal
relative purity and generalized unentanglement still remains an
interesting open question in the general convex cone framework.  We
refer the reader to the above-mentioned papers for an extended
discussion of this issue, as well as for the construction of
appropriate {\em mixed-state} GE measures.  In what follows, we focus
on developing concrete intuition about GE based on several
illustrative examples.

\section{Generalized entanglement by example}

\subsection{Entanglement with respect to local observables}

Assume that the system $S$ is composed of $n$ distinguishable
subsystems, corresponding to a Hilbert-space factorization of the form
\begin{equation}
{\cal H}\simeq {\cal H}_1\otimes \ldots \otimes {\cal H}_n,\label{tps}
\end{equation}
\noindent 
where for simplicity we take the factors to be isodimensional,
dim$({\cal H}_\ell)=d_0$ for all $\ell$, $d_0^n=d$.

\subsubsection{The Lie-algebraic setting}

Contact with the standard multipartite entanglement framework may be
established by realizing that full {\em local accessibility} of
individual subsystem states identifies the set of {\em arbitrary local
observables} as the physically distinguished set.  Let
\begin{equation}
\Omega_{loc}\equiv \mathfrak{h}_{loc}=\mathfrak{su}(d_0)_1
\oplus\ldots \oplus \mathfrak{su}(d_0)_n
\label{hloc}
\end{equation} 
denote the relevant (irreducible) Lie algebra of traceless local
observables, generalizing the bipartite qubit case of
Eq.~(\ref{locobsp}).  For each $\mathfrak{su}(d_0)_\ell $, an
orthonormal (in the trace inner product) basis $\{x_{\alpha} \}$,
$\alpha=1, \ldots, d_0^2-1$, may be constructed in analogy to
(normalized) Pauli operators.  Thus $\mbox{Tr}(x_\alpha x_\beta) =
\delta_{\alpha, \beta}$.  An overall orthonormal basis for
$\mathfrak{h}_{loc}$ is then obtained by extending each
$x^\ell_{\alpha}$ to act non-trivially only on subsystem $\ell$, e.g.
$$x^{(1)}_{\alpha}= x_\alpha \otimes \openone^{(2)} \otimes\ldots\otimes
\openone^{(n)},\;\; \openone^{(\ell)}=\openone/\sqrt{d_0}\,.$$
\noindent
While a formal proof of the equivalence between standard entanglement
and GE relative to arbitrary local observables is given in
\cite{BarnumPRA03}, the basic step follows from identifying reduced
local states with lists of expectations of $x^\ell_{\alpha}$: Because
the latter completely determine reduced density operators (in the
usual partial-trace sense), and a pure state $|\psi\rangle \in {\cal
H}$ is entangled (in the standard sense) if and only if all its
reduced density operators remain pure, the two notions coincide on
pure states.  By convexity, they can also be shown to coincide on
mixed states.  Consistently, GCSs of $\mathfrak{h}_{loc}$ may be
associated to orbits of a reference state like $|0\ldots0\rangle$
under the group of local (special) unitary transformations,
SU$(d_0)_1\otimes \ldots \otimes \mbox{SU}(d_0)_n$.

The connection with the ordinary entanglement framework may be further
quantitatively appreciated by relating the local purity
$P_\mathfrak{h}\equiv P_{loc}$ to the conventional subsystem purities
determined by reduced subsystem states.  One finds \cite{ViolaCM05}
\begin{equation}
P_{loc}(|\psi\rangle) = \frac{d_0}{d_0-1}\Big(
\frac{1}{n}\sum_{\ell=1}^n \mbox{Tr}(\rho_\ell^2) -\frac{1}{d_0}
\Big)\,,
\label{Ploc}
\end{equation}
where $\rho_l$ denotes as usual the reduced density operator of
subsystem $\ell$.  Thus, $P_{loc}$ is proportional to the {\em
average} subsystem purity.  For the special case of qubits ($d_0=2$),
the local GE as quantified by $P_{loc}$ is additionally simply related
\cite{SommaPRA04,Brennen} to a measure of {\em global multipartite
entanglement} $Q$ originally proposed in \cite{MW},
$P_{loc}(|\psi\rangle) =1 - Q (|\psi\rangle).$

\subsubsection{A three-qubit case study}

It is essential to realize that in order for the equivalence between
multipartite subsystem entanglement and GE to hold, {\em all} and {\em
only} local observables must be distinguished.  A concrete example may
serve to further illustrate these points. Let $S$ consists of three
qubits.  The local algebra is given by
\begin{equation}
\Omega_1 =\mathfrak{su}(2)_1 \oplus\mathfrak{su}
(2)_2 \oplus\mathfrak{su}(2)_3\,,
\label{s1}
\end{equation}
with corresponding local purity given by
$$ P_1(|\psi\rangle) ={1\over 3} \sum_{\ell=1}^3 \sum_{\alpha=x,y,z}
\langle \sigma^{(\ell)}_\alpha \rangle^2 \,.$$
\noindent 
This distinguishes four classes of states with different multipartite
(and GE) properties: $P_1(|\psi\rangle) =1$ on arbitrary product
states, which are thus unentangled; $P_1(|\psi\rangle) =1/3$ on
so-called bi-separable states -- joint states e.g. of the form
$$ |{\tt B}_{12}\rangle = |{\tt Bell}\rangle_{12} \otimes |\phi\rangle_3
\,,$$
\noindent
and similar for states $|{\tt B}_{13}\rangle, |{\tt B}_{23}\rangle$
where qubit $2, 3$ are factored out, respectively. Next,
$P_1(|\psi\rangle) =1/9$ for $|\psi\rangle$ belonging to the so-called
{\tt W}-class, e.g.
$$ |{\tt W} \rangle = {1\over \sqrt{3}}\left( 
|001 \rangle + |010 \rangle + |100 \rangle \right)\,,$$
\noindent
and $P_1(|\psi\rangle) =0$ on the Greenberger-Horne-Zeilinger ({\tt
GHZ}) class,
$$ |{\tt GHZ} \rangle = {1\over \sqrt{2}}\left( |000 \rangle + |111
\rangle \right) \,,$$
\noindent
identifying the latter as maximally entangled with respect to local 
observables. 

Suppose, however, that ``local'' resources are redefined, so that
arbitrary unitary operations on the first pair are distinguished --
that is, we effectively replace $\Omega_1$ with 
\begin{equation}
\Omega_2 =\mathfrak{su}(4)_{12} \oplus\mathfrak{su}(2)_3\,,
\label{s2}
\end{equation}
and, correspondingly, we compute the new ``bi-local'' purity
$P_2(|\psi\rangle)$ by using an appropriate basis of $\Omega_2$ --
including $6$ bilinear Pauli operators in addition to the $9$ of
$\Omega_1$.  How does the above classification change?  Were qubit 3
not present, then clearly any pure state in ${\cal H}_1 \otimes {\cal
H}_2$ would be disentangled relative to the full algebra
$\mathfrak{su}(4)_{12}$.  With qubit 3 included, one still expects any
internal structure within qubits 1 and 2 to be irrelevant, as long as
the pair reduced state remains pure.  This intuition is reflected by
behavior of $P_2(|\psi\rangle)$, which now attains its maximum on both
product states as before {\em and} the bi-separable class ${\tt
B}_{12}$ -- which thus becomes extremal.  Both ${\tt B}_{23}$ and
${\tt B}_{13}$ become maximally generalized entangled relative to this
observer, along with $|{\tt GHZ} \rangle$ ($P_2=1/3$), whereas the
$|{\tt W} \rangle$ class shows intermediate GE with $P_2(|{\tt W}
\rangle)=11/27$.

Other interesting scenarios may be conceived, which resemble in part
ordinary entanglement -- in the sense that partial accessibility of
subsystems in ${\cal H}$ is retained -- yet differ in the
identification of actual ``local'' degrees of freedom -- as
observables spaces on different factors overlap.  Suppose, for
instance, that qubits 1, 2, 3 are spatially separated and arranged on
a line, or reside at the vertexes of a triangle, with physical
transformations generated by nearest-neighbor two-body couplings in
the first case, or arbitrary two-body couplings in the second one.
Then the resulting operational constraints are naturally captured by
distinguished observable sets of the form
\begin{eqnarray}
\Omega_3 =\mbox{span}_{\mathbb R}\{ \sigma^{(1)}_a \otimes
\sigma^{(2)}_b \otimes \openone^{(3)}, \openone^{(1)} \otimes
\sigma^{(2)}_b \otimes \sigma^{(3)}_c \}\,,
\label{s3} 
\end{eqnarray}
or, respectively, 
\begin{eqnarray}
\Omega_4 =\mbox{span}_{\mathbb R}\{ &\hspace*{-1.5mm} \sigma^{(1)}_a
\hspace*{-1.5mm}& \otimes \sigma^{(2)}_b \otimes \openone^{(3)},
\openone^{(1)} \otimes \sigma^{(2)}_b \otimes \sigma^{(3)}_c,
\nonumber \\ & \hspace*{-1.5mm}\sigma^{(1)}_a \hspace*{-1.5mm}&
\otimes \openone^{(2)} \otimes \sigma^{(3)}_c\}\,.
\label{s4} 
\end{eqnarray}
A formal analysis of GE relative to the above distinguished sets is
technically more difficult as neither of $\Omega_{3,4}$ is a Lie
algebra.  While such an analysis is beyond our current purposes,
following the GE behavior of different classes of pure states as
observables are added through the progression $\Omega_1 \rightarrow
\Omega_2 \rightarrow \Omega_3 \rightarrow \Omega_4$ clearly serves to
illustrate how GE is directly observer-dependent irrespective of
whether a direct correspondence is possible between the concept of
``locality'' and ``locally accessible'' subsystem degrees of freedom:
note, in particular, that relative to $\Omega_4$ every bi-separable
pure state of the three qubits becomes extremal -- thus generalized
unentangled, consistently with physical intuition.

\subsubsection{The convex cones setting; GE in Popescu-Rohrlich boxes}

Although the convex cones setting is more general than quantum
mechanics, in this setting one may still define a natural notion of
one system's being a subsystem of another, and of a system's being a
tensor product of two systems.  In this case, however, the notion of
tensor product is not unique.  For our purposes, we will (as in
\cite{BBLW06}) allow as a tensor product of cones $C_1 \subset V_2$
and $C_2 \subset V_2$ {\em any} cone $\Gamma$ in $V_1 \otimes V_2$,
with base $\hat{\Gamma}$ containing the {\em separable} tensor product
$\hat{C_1} \otimes_{sep} \hat{C_2}$, and contained in the {\em
maximal} tensor product $\hat{C}_1 \otimes_{max} \hat{C}_2$.  Both the
separable and maximal tensor product constructions admit a simple
interpretation.  The separable tensor product is the convex hull of
the products $x \otimes y$, $x \in \hat{C}_2$, $y \in \hat{C}_2$.
These product states are just the natural generalization of classical
product distributions and quantum product states: for any pair of
observables on $C_1$ and $C_2$, the joint probability given by such
states, for outcome-pairs, factorizes as a product of marginal
distributions for the two systems.  The maximal tensor product is just
$(C_1^* \otimes_{sep} C_2^*)^*$, i.e. it is the set of unnormalized
states that are nonnegative on all ``product outcomes''.  States in
the maximal tensor product are precisely those that do not allow
signalling.  This general notion of tensor product appeared in
\cite{Namioka-Phelps}, and (in a slightly different but essentially
equivalent formalism of test spaces) the no-signalling construction of
the maximal tensor product is done in
\cite{Foulis-Randall,Randall-Foulis}.  A review of the latter (with
additional results on certain tensor products of two {\em quantum}
systems) can be found in \cite{BFRW05}, along with many references to
related work, and the notion of ``system combination'' independently
developed in \cite{Barrett05a} should also be noted as closely
related.

One can specialize our GE notion (and, correspondingly, generalize the
notion of bipartite quantum entanglement) to such a tensor product, by
letting $D \subset V_1 \otimes V_2$ be the cone generated by
restricting states in $\Gamma$ to the space of effects spanned by $\{x
\otimes \lambda, \lambda \otimes y\,|\, x \in C^*_1, y \in C_2^*\}$.
There is a natural map $\pi$ satisfying the conditions stated in the
definition of GE.  For any state $\omega$ in $\Gamma$, its marginal
states $\omega_1, \omega_2$ on system $1$ or $2$ may be defined as the
restrictions of $\pi$ to the spaces of observables on system $1$ or
$2$ respectively, i.e., to $\{x \otimes \lambda\,|\, x \in C_1\}$ or
$\{\lambda \otimes y\,| \, y \in C_2^*\}$.  The reduced states
$\pi(\omega) \in D$ are thus just pairs $(\omega_1, \omega_2)$ of
marginal states, and {\em whatever} tensor product between the maximal
and the separable is chosen, it turns out that, relative to $D$ and
$\pi$, the unentangled bipartite states in that tensor product are
just the ones in the {\em separable} tensor product.  That is, pure
unentangled states are just products of pure (extremal) states, and
general unentangled states are convex combinations of these.  This
follows from the fact (see e.g. Lemma 3 of \cite{BBLW06}, though this
is almost certainly not its first appearance) that if either marginal
state of some state in a bipartite tensor product as we have defined
it, is pure, the bipartite state is a product state.

As a concrete non-quantum, non-classical example of the distinction
between product and entangled pure states, consider the extremal
states of a pair of two-output, two-input {\em Popescu-Rohrlich (PR)
boxes}.  In terms of convex sets, a single such box may be viewed as
having a state space that is polyhedral with square base.  This can be
visualized in ${\mathbb R}^3$, although it is in some ways more
natural to view as a non-generating cone in ${\mathbb R}^4$, as the
interpretation below will clarify.  Suppose we have chosen an
orthonormal basis, and let the distinguished ($\lambda(x)=1$) square
base lie in the plane perpendicular to a line between the center of
the square base and the origin.  The cone of unnormalized measurement
outcomes (the dual cone to this) is thus also a square polyhedral
cone: if we visualize it in the same space and represent evaluation of
functionals by the Euclidean inner product in this space, it will be
rotated by $\pi/4$, relative to the primal cone, around the line
through the center of the square and the origin (and also uniformly
dilated or contracted relative to it).  No matter how we affinely
embed the primal cone in ${\mathbb R}^3$, the dual cone will not
coincide with it, so this cone does not enjoy the important property
of {\em self-duality}, but it is still isomorphic to its dual.

The interpretation of the states is in terms of two alternative
measurements, each having two possible outcomes, the measurements
labelled in the usual PR box formalism by an ``input'' bit specifying
which measurement is to be done, and the outcome of each measurement
by an ``output'' bit -- so that, taken together, these two
measurements have a total of four possible measurement outcomes, each
corresponding to a pair of an input and an output bit.  Pictorially,
such outcomes correspond to points on the four extremal rays of the
dual cones.  View the square state space as embedded in ${\mathbb
R}^2$, so that the vertexes are $(0,0)$, $(0,1)$, $(1,1)$, $(1,0)$ as
we go clockwise around the square base, and arbitrary states
correspond to points $(p_0, p_1) \in [0,1] \times [0,1]$.  Interpret
$p_0$ as the probability of measurement $0$ yielding the result $0$,
$(1-p_0)$ as the probability of it yielding the result $1$, $p_1$ as
the probability of measurement $1$ yielding $0$, and $(1-p_1)$ as the
probability of measurement $1$ yielding $1$.  Thus, the two-parameter
representation of the state can be viewed as the result of an affine
embedding of the subset of the four-parameter states $(p_{0|0},
p_{1|0}, p_{0|1}, p_{1|1})$, satisfying the normalization constraints
$p_{0|0} + p_{1|0} = 1$, $p_{0|1} + p_{1|1} = 1$, into ${\mathbb
R}^2$.  Similarly the cone of unnormalized states in ${\mathbb R}^3$
can be viewed as an isomorphic affine image (embedding) of the
three-dimensional subspace of the four parameter states in the octant
${\mathbb R}^4_+$ satisfying the constraint $p_{0|0} + p_{1|0} =
p_{0|1} + p_{1|1}$.

By a similar construction, analogous boxes can be defined for an
arbitrary finite number of ``inputs'' $N$ (corresponding to
alternative measurements), and numbers $M_i, i \in \{1,...,N\}$ of
possible outputs for each measurement (usually taken to be a number
$M$ independent of $i$).  Such sets of alternative measurements are
also known in the quantum logic literature as {\em semi-classical test
spaces} \cite{Wilce2000}.

The state-space of a pair of PR boxes is just the maximal tensor
product of the state spaces of two PR boxes: it was introduced
\cite{PR94} in the PR box literature as the set of all possible states
of correlations between the outcomes of two PR boxes that do not
permit signalling.  This state space is a polytope; it has been
studied in \cite{BLMPPR05} and, in particular, its extremal points for
the two-input, $d$-output case have been explicitly described.  For
our case, $d=2$, there are two types of extremal states.  They are
best understood by exhibiting states as $4 \times 4$ matrices of
probabilities $p_{ij|kl}$, where $ij$ is the pair of Alice's outcome
label, Bob's outcome label (``outputs''), and $kl$ is the pair of
Alice's measurement, Bob's measurement (``inputs'').  Thus the matrix
has a natural $2 \times 2$ block structure, the $ij$-th block being
the $2 \times 2$ matrix for the probabilities of the four possible
Alice/Bob outcome-pairs when Alice does measurement $k$, Bob
measurement $l$.  That is, explicitly, 
\begin{equation}
\left(
\begin{array}{ll|ll}
p_{00|00} & p_{01|00} & p_{00|01} & p_{01|01} \\
p_{10|00} & p_{11|00} & p_{10|01} & p_{11|01} \\
\hline 
p_{00|10} & p_{01|10} & p_{00|11} & p_{01|11} \\
p_{10|10} & p_{11|10} & p_{10|11} & p_{11|11} 
\end{array}
\right) \: .
\end{equation}

The standard normalization condition is that the probabilities within
each block sum to $1$.  Bob's marginal state is the vector of column
sums of the upper half of the matrix, interpreted as $(p_{0|0},
p_{1|0}, p_{0|1}, p_{1|1})$; the condition that Alice cannot signal to
Bob translates into the fact that this is equal to the vector of
column sums of the lower half of the matrix (i.e., Bob's probabilities
are independent of Alice's choice of measurement).  Similarly, Alice's
marginals are the row sums of the left half of the matrix, equal by
no-signalling to those of the right half.  The fact that these are
linear equality constraints (and the positivity constraint on
probabilities a linear inequality constraint) is what makes the
no-signalling state-space a polytope and, consequently, makes the cone
of unnormalized states a polyhedral cone.

The extremal points were found and classified in \cite{BLMPPR05} and,
as mentioned, are of two types.  One is represented by the following
state:
\begin{equation}
\left(
\begin{array}{ll|ll}
1 & 0 & 1 & 0 \\
0 & 0 & 0 & 0 \\
\hline 
1 & 0 & 1 & 0 \\
0 & 0 & 0 & 0 
\end{array}
\right) \: .
\end{equation}
Alice's reduced state is just the marginal state $(1,0,1,0)$, as is
Bob's; these are both extremal states, so this is generalized
unentangled by our definition: indeed, it can also easily be seen to
be a product state, $p_{ij|kl} = p^A_{i|k} p^B_{j|l}$.  There are $16$
representatives of this class, representing products of each of the
four local extremal states for Alice with each of four local extremal
states for Bob.

The other class consists of eight entangled extremal states.  These
are all locally equivalent to the representative:
\begin{equation}
\left(
\begin{array}{ll|ll}
\half  & 0 & \half & 0 \\
0 & \half & 0 & \half 
\vspace*{0.75pt} \\
\hline 
\half & 0 & 0 & \half \\
0 & \half & \half & 0
\end{array}
\right) \:. 
\end{equation}
In this case, both Alice and Bob's marginals are the mixed state
$(\half, \half, \half, \half)$, showing (given that these are extremal
overall, as proved in \cite{BLMPPR05}) that this state is generalized
entangled by our definition.  Of course, the fact that this is
entangled in the sense that it is a pure non-product state was already
observed in \cite{BLMPPR05}.  As noted there, the other 15 pure
product states, and the other 7 pure entangled states, can be obtained
from the above two states by local transformations consisting of
relabeling the measurements and the outcomes.  The entangled states,
for example, may be described via the shorthand $A$ (for
anti-correlated) for the $2 \times 2$ matrix that looks like
$\sigma_x/2$, and $C$ (for correlated) for the $2 \times 2$ matrix
$\openone/2$.  In this way, the above entangled state reads:
\begin{equation}
\left(
\begin{array}{ll}
C  &  C \\
C & A 
\end{array}
\right) \;, 
\end{equation}
and all $8$ entangled states may be obtained by noting that relabeling
the measurements just interchanges rows and/or columns -- allowing us,
with a few such relabellings, to place the $A$ in any of the four
places, whereas relabeling the outcomes of one of Alice's or Bob's
measurements interchanges $C$ and $A$ in the row (for Alice) or column
(for Bob) corresponding to that measurement -- allowing us access to
the four states having three $A$'s and one $C$.

Ref.~\cite{BLMPPR05} also describes the single local equivalence class
of entangled pure states of a pair of two-measurement, $d$-outcome
boxes, and some of the $44$ local equivalence classes of entangled
pure states of three two-measurement, two-outcome boxes, along with
many other interesting results, for example on local interconversion
of various combinations of states of PR boxes.  Further properties of
operational theories based on PR boxes are investigated in
\cite{Barrett05a}.

\subsection{Entanglement without locality...}

The extent to which the GE notion genuinely enlarges the standard
subsystem-based framework may be further appreciated in situations
where ${\cal H}$ is intrinsically indecomposable -- thus no
factorization of ${\cal H}$ exists and conventional entanglement is
not directly applicable.  A physically motivated example is offered by
a quantum spin-$J$ system (take for definiteness $J\in {\mathbb N}$),
whose $d=(2J+1)$-dimensional state space carries an irreducible
representation of $\mathfrak{su}(2)$, with angular momentum generators
$J_\alpha$ satisfying commutation rules $[J_\alpha, J_\beta] =
2i\epsilon_{\alpha\beta\gamma} J_\gamma$,
$\epsilon_{\alpha\beta\gamma}$ denoting the completely antisymmetric
tensor.  Because the GE notion rests only on convex properties of sets
of quantum states and observables, the definition of GE is still
applicable as soon as distinguished observable sets are specified. In
particular, the full algebra of traceless observables $\mathfrak{g}$
is isomorphic to $\mathfrak{g}\simeq \mathfrak{su}(d)$: clearly,
expectations in $\mathfrak{g}$ completely determine an arbitrary pure
state in ${\cal H}$ -- accordingly, every pure state is (extremal)
unentangled relative to $\mathfrak{g}$.

Imagine, however, that only expectations of observables in the
``local'' sub-algebra $\mathfrak{h}=\mathfrak{su}(2)$ are available.
In this case, $\mathfrak{h}$-reduced states may be identified with
three-dimensional vectors of expectation values $\langle
J_\alpha\rangle$, $\alpha=x,y,z$, which form a ball of radius $J$ in
${\mathbb R}^3$.  Using the notation ${\bf J}$ for the vector $(J_x,
J_y, J_z)$, these are vectors $\langle {\bf J} \rangle$.  In addition,
${\bf J}^2 = {\bf J}\cdot {\bf J} = J_x^2 +J_y^2 + J_z^2$.  The
extremal states corresponds geometrically to points on the surface of
such a ball, that is to the GCSs of SU(2) -- also called {\em spin
coherent states} (SCSs) \cite{Pere} -- which are characterized by
maximal spin projection along a given component,
$$ ({\bf n}\cdot {\bf J})|\psi\rangle_{{\tt SCS}}= J
|\psi\rangle_{{\tt SCS}},\; |{\bf n}|=1 \,.$$
\noindent
For any given choice of spin direction, e.g. ${\bf n}=\hat{z}$, let
$\{|J, J\rangle, $$|J, J-1\rangle, \ldots, |J,0\rangle,\ldots, |J,
-J+1\rangle, |J,-J\rangle\}$ be an orthonormal basis of ${\cal H}$
consisting of joint ${\bf J}^2, J_z$ eigenstates.  Then states $|J,\pm
J\rangle$ lie on the surface and therefore are generalized
unentangled, whereas the remaining states lie on the inside, so are
not extremal: in particular, the state $|J,0\rangle$ lies on the
middle of the sphere and is {\em maximally entangled} relative to
$\mathfrak{su}(2)$. Mathematically, this is reflected in minimal
$\mathfrak{su}(2)$-purity,
$$ P_{\mathfrak{su}(2)}(|J,0\rangle) = \frac{1}{J^2}
\sum_{\alpha=x,y,z} \langle J,0| J_\alpha| J,0\rangle^2 =0\,.$$
\noindent
Physically, the presence of GE captures the fact that no ``local''
unitary operation (no rotation in SU(2)) is able to connect such a
state with extremal states on the surface: achieving that requires
``entangling'' operations generated by Hamiltonians in the full
$\mathfrak{g}$.

Note that because $\mathfrak{h}$ is irreducible, the
$\mathfrak{h}$-purity coincides, up to an additive constant, with the
quantity
$$(\Delta {\cal I})^2 =\sum_\alpha\left[ \langle J^2_\alpha\rangle -
\langle J_\alpha \rangle^2 \right]= J(J+1) - J^2 P_{\mathfrak{su}(2)}
\,,$$
\noindent 
which measures the so-called {\em invariant uncertainty} of SU(2) and
is minimized by GCSs \cite{Pere,Sergio}.  From this point of view,
generalized unentangled (entangled) states may thus be seen as
maximally close to (remote from) ``classical reality'' -- as measured
by the corresponding minimal (maximal) amount of quantum fluctuations
\cite{Klyachko, Klyachko06}.

\subsection{Separability without entanglement...}

A further element of distinction between conventional and generalized
entanglement arises in situations where a local structure exists in
${\cal H}$ as in Eq.~(\ref{tps}), but operational capabilities are
restricted to a {\em proper subset of local observables}.  

Consider, for instance, two spin-$J$ particles, with associated
bipartite Hilbert space ${\cal H}= {\cal H}_1 \otimes {\cal H}_2$,
$d_0=2J+1$, and full observable Lie algebra $\mathfrak{g}\simeq
\mathfrak{su}(d^2)$.  Arbitrary (subsystem)-local observables
correspond to $\mathfrak{h}_{loc}\simeq
\mathfrak{su}(d)_1\oplus\mathfrak{su}(d)_2$. If, as in the above
example, only observables linear in the basic angular momentum
generators are distinguished for each subsystem, then the accessible
``local'' sub-algebra is
$$\mathfrak{h'}_{loc} = \mathfrak{su}(2)_1\oplus\mathfrak{su}(2)_2
\subset \mathfrak{h}_{loc}\,.$$ 
\noindent
The set of generalized unentangled states relative to
$\mathfrak{h'}_{loc}$ consists of states of the form $|{\tt
SCS}\rangle_1 \otimes |{\tt SCS}\rangle_2$.  However, other {\em
tensor product} states like $|J,0\rangle_1 \otimes |J,0\rangle_2$ are
maximally entangled relative to the local sub-algebra: to fully
``resolve'' the product nature of such states requires access to
expectations of arbitrary observables in $\mathfrak{h}_{loc}$.

\subsection{Fermionic entanglement}

As a last example, we reconsider the simplest instance of
entanglement in systems of indistinguishable fermions, briefly
mentioned at the end of Sect. II.

Because the requirement of complete antisymmetry of the joint state
vector under particle exchange is automatically incorporated,
identical (spinless) fermions are most naturally described in a
second-quantized language based on canonical fermion operators.
Consider two identical fermions, each of which may occupy one of two
modes~\cite{Ortiz}.  For each mode $j=1,2$, let $c^\dagger_j$, $c_j$
denote creation and annihilation operators respectively, obeying the
following anti-commutation rules:
\begin{eqnarray*}
&& [c_i, c_j]_+ = [c^\dagger_i, c^\dagger_j]_+ = 0\,,\\
&& [c^\dagger_i, c_j]_+ =\delta_{ij} \,,
\end{eqnarray*}
where $[\,,\,]_+$ denotes the anti-commutator. Let in addition $|{\tt
vac}\rangle$ denote a reference ``vacuum'' state containing no
fermions.  For instance, we may associate mode 1 with {Alice} and mode
2 with {Bob}, who reside at spatially separated locations.  Then a
state where a fermion is created at {Alice}'s site corresponds to
$|c^\dagger_1\rangle \equiv c^\dagger_1|{\tt vac} \rangle$, and so on.

What kind of entanglement is physically meaningful in fermionic
states? Insight into this question may be sought by changing the
description into a more familiar spin language, by exploiting the
Jordan-Wigner mapping between the above fermionic operators and the
Pauli algebra,
\begin{eqnarray*}
S^{(1)}_+ = c_1^\dagger,\; && S^{(1)}_- =(S^{(1)}_+)^\dagger \,,\\
S^{(2)}_+ = (1-2 n_1) c_2^\dagger, \; && S^{(2)}_-
=(S^{(2)}_+)^\dagger,
\end{eqnarray*}
where $\hat{n}_1=c_1^\dagger c_1$ is the fermion number operator for
mode 1, and the operators $S^{(j)}_\pm =(\sigma_x^{(j)} \pm i
\sigma_y^{(j)})/2$ so constructed obey $\mathfrak{su}(2)$ commutation
rules. Then the following correspondence may be established between
states in the usual spin-$\frac{1}{2}$ basis $\{ |0\rangle,
|1\rangle\}$ and fermionic states
\begin{eqnarray}
&& |00\rangle \leftrightarrow |{\tt vac}\rangle \nonumber \,, \\ &&
|01\rangle \leftrightarrow c_1^\dagger|{\tt vac}\rangle \nonumber \,,\\ &&
|10\rangle \leftrightarrow c_2^\dagger|{\tt vac}\rangle \nonumber \,, \\ &&
|11\rangle \leftrightarrow c_1^\dagger c_2^\dagger|{\tt vac}\rangle\,.
\label{onep}
\end{eqnarray}
Presumably, we would want none of these states (containing,
respectively, zero, one, one, and two fermions at different locations)
to be ``entangled'' by any reasonable definition.  What about the set
of Bell states given in Eq.~(\ref{Bell})?  By applying the
Jordan-Wigner mapping, these can be rewritten as follows:
\begin{eqnarray*}
|\Phi^\pm\rangle =\frac{1}{\sqrt{2}} (|01\rangle \pm |10\rangle) 
& \leftrightarrow & \frac{1}{\sqrt{2}}(c_1^\dagger \pm
 c_2^\dagger) |{\tt vac}\rangle \,,\\
|\Psi^\pm\rangle =\frac{1}{\sqrt{2}}(|00\rangle \pm |11\rangle)  
& \leftrightarrow & \frac{1}{\sqrt{2}}(\openone \pm
 c_1^\dagger c_2^\dagger ) |{\tt vac}\rangle = \\ 
& = & \frac{1}{\sqrt{2}}(|{\tt vac}\rangle \pm 
c_1^\dagger c_2^\dagger  |{\tt vac}\rangle)\,.
\end{eqnarray*}
Note that states in the upper line are still states with a {\em fixed}
number of particles (one particle), which are described in general by
so-called ``Slater determinants'' in the condensed-matter terminology.
States in the bottom line, in contrast, are linear combinations of
states with zero and two fermions that is, linear combinations of
Slater determinants with {\em different} particle number.  

Suppose we consider an arbitrary physical scenario or process where
{\em no} change of the fermion number in a state can occur.
Then admissible physical observables for fermions must commute with
the total fermion operator,
$$ \hat{N}=\hat{n}_1 + \hat{n}_2 \rightarrow \frac{\sigma_z^{(1)}}{2}
+ \frac{\sigma_z^{(2)}}{2} -\openone = S_z -\openone \,,$$
\noindent
which recovers precisely the conservation law discussed in
Sect. II. In fact, bilinear fermion operators of the form $\{
c^\dagger_i c_j, 1\leq i,j \leq 2\}$ may be seen to satisfy
$\mathfrak{u}(2)$-commutation rules \cite{SommaPRA04}, which makes the
latter a natural candidate for a distinguished fermionic observable
set.  A larger fermionic algebra arises if arbitrary bilinear fermion
operators are included (e.g. operators of the form $c^\dagger_i
c^\dagger_j$), leading to $\mathfrak{so}(4) \supset \mathfrak{u}(2)$.
Written in the fermionic language, the $\mathfrak{u}(2)$ Lie algebra
given in Eq.~(\ref{u2}) becomes
\begin{eqnarray*}
\mathfrak{u}(2)=\mbox{span}_{\mathbb R}\{ \hat{n}_1-1/2,
\hat{n}_2-1/2, (c^\dagger_1 c_2+c^\dagger_2 c_1)/\sqrt{2},&& \\ i
(c^\dagger_1 c_2 - c^\dagger_2 c_1)/\sqrt{2} \}\,.&&
\end{eqnarray*}
The corresponding $\mathfrak{u}(2)$-purity, $P_{\mathfrak{u}(2)}$,
(computed either in the fermionic or in the spin language) attains its
maximum on both the number eigenstates of Eq.~(\ref{onep}) and the two
Bell states $|\Phi^\pm\rangle$.  Thus, the latter states are certainly
``mode-entangled'' with respect to the local spin algebra
$\Omega_{loc}$, but generalized unentangled relative to the fermionic
algebra, consistent with the original intuition about their
one-particle nature.  In contrast, states $|\Psi^\pm\rangle$ have zero
$\mathfrak{u}(2)$-purity, thus they are maximally entangled both in
the conventional spin sense {\em and relative to the $\mathfrak{u}(2)$
observer}.  That is, states $|\Psi^\pm\rangle$ contain genuine
fermionic entanglement -- irrespective of the operator language used.
Physically, this expresses the fact that a linear combination of
Slater determinants with different fermion numbers {\em cannot} be
distinguished from a mixture by relying solely on expectations in
$\mathfrak{u}(2)$: indeed, distinguishing would involve expectations
of operators (like e.g. $\sigma_x^{(1)}\otimes \sigma_x^{(2)}$ in
$\mathfrak{so}(4)$) which have non-zero matrix elements between states
of different fermion number, hence require the above conservation law
to be broken \cite{obs}.

\section{Generalized entanglement: Applications and implications 
(so far...)}

The GE notion has proved so far a powerful unifying framework for
linking entanglement properties to various physical,
information-theoretic, and conceptual aspects of ``complexity'' and
``classicality'' emerging in a variety of scenarios.

\subsection{Complexity implications}

Following one of the original motivations underlying GE
\cite{SommaPRA04,Ortiz}, purity measures associated to appropriate
observable subspaces are providing novel diagnostic tools for
characterizing the ``correlations'' present in eigenstates of
interacting quantum many-body Hamiltonians -- allowing natural contact
to be established with state-complexity notions like the {\em number
of principal components} or the {\em inverse participation ratio},
borrowed from quantum statistical physics \cite{Winton}.  Two
situations are especially relevant in this regard, and have been the
subject of intense investigation recently (see
e.g. \cite{osborne,vidal,fazio,casati} and references therein).  On
one hand, the correlations present in the {\em ground-state} of a
many-body system may undergo a structural change as some parameters in
the Hamiltonian are changed at zero temperature across ``critical''
values -- giving rise to so-called ``quantum phase transitions'' in
the limit of infinite-system size.  In this case, natural
Lie-algebraic GE measures constructed from fermionic and/or spin
operators have made it possible to successfully identify and
characterize the ensuing critical behavior in a large class of
transitions induced by a spontaneous symmetry breaking
\cite{SommaPRA04,Remark}.  On the other hand, structural changes may
also take place for {\em any} typical many-body state if the addition
of a perturbation or disorder to the original Hamiltonian causes a
crossover from a regular, ``integrable'' regime into non-integrability
and so-called ``quantum chaos''.  The onset of chaos may in turn
manifest itself at both the static level -- in terms, for instance, of
different eigenvector statistics as described by so-called ``random
matrix theory'' \cite{rmt} -- and at the dynamic level -- in terms of
different behavior of quantum ``fidelity'' as a function of time
\cite{peres,CPJ05,prosen}, or hypersensitivity to perturbations
\cite{SchackCaves, SBCS06}.  Investigations in paradigmatic systems
such as disordered quantum spin lattices \cite{Montangero} and chaotic
quantum maps \cite{Weinstein} have given clear indication so far that
GE and GE generation with respect to appropriate observables can serve
as reliable indicators for detecting and probing quantum chaos.

Suggestively, the GE notion has also recently shed light on the
conditions needed to unlock the full power of quantum computational
models as compared to purely classical ones.  For a large class of
``Lie-algebraic quantum computations'', which are specified through
controllable interactions and measurable observables in a Lie algebra
$\mathfrak{g}$ and initialization of the system in a GCS of
$\mathfrak{g}$, the results in \cite{SommaPRL06} demonstrate how GE is
required for genuinely stronger-than-classical computational models to
emerge.

\subsection{Classicality implications}

If we consider a quantum system with the distinguished set of
observables consisting of all those observables diagonal in some {\em
fixed} basis (a special case of the associative-algebraic setting, in
fact a case where the associative algebra is commutative), the
generalized unentangled states are precisely the set of density
matrices diagonal in that basis.  Often the selection of such a
distinguished basis is viewed as the selection of a distinguished
``classical'' set of states because dynamics that preserve diagonality
of density matrices in such a basis, together with measurements of
observables diagonal in the same basis, are equivalent to a classical
theory on a number of classical states equal to the dimension.  This
notion also generalizes to the convex cones setting as follows.  In
\cite{BBLW06}, clonable sets of states were shown to be jointly
distinguishable by a single measurement, and vice versa.  Also,
broadcastable sets of states were shown to belong to the convex hull
of such a set of jointly distinguishable states (which we term a
``simplex generated by distinguishable states'', or SGDS).  Indeed,
for any map $B$ from a convex set $\Omega$ to a tensor product $\Omega
\otimes \Omega$, the set of states broadcast by $B$ was shown to be
{\em precisely} such a simplex (though for some maps it will be the
empty set, viewed as a degenerate case of such a simplex).  A
(finitely generated) simplex is the convex hull of $n$ or fewer points
in ${\mathbb R}^n$, for some finite $n$; the theory whose set of
normalized states is an $n$-simplex in ${\mathbb R}^n$ and whose cone
of measurements is the dual of this (which is also based on an
$n$-simplex), is classical.  As the projectors $| i \rangle \langle
i|$ onto the states of some orthonormal basis form a $d$-dimensional
simplex in the $d^2$-dimensional space of Hermitian operators on a
$d$-dimensional quantum system, and as they are distinguishable by a
single measurement, they are a special case of an SGDS as defined
above.  For an SGDS generated by states $\omega_i \in C$, if 
$a_i \in C^*$ such that $\sum_i a_i = \lambda$ is a measurement
distinguishing the states $\omega_i$, then we may take as our cone $D$
the cone of reduced states, defined as the restrictions of states in
$C$ to the space of functionals in the span of the $a_i$. Then states
in the SGDS are precisely the generalized unentangled states relative
to the observables generated by the $a_i$: that is, classical states
with respect to this set of distinguished observables.

The problem of identifying classicality aspects of generalized
unentangled states has also been tackled recently by making explicit
contact with the open quantum system theory and the decoherence
program \cite{Zurek}.  Within a Lie-algebraic formulation of Markovian
quantum dynamics, in particular, GCSs are found to minimize an
invariant uncertainty which is closely related to their quantum Fisher
information content, and they are seen to emerge as most predictable
``ein-selected'' pointer states under appropriate conditions on the
interaction between the system and its surrounding environment --
generalizing the characterization of canonical (harmonic-oscillator)
coherent states as most stable (hence most classical) states in the
presence of decoherence \cite{Sergio}.

\subsection{Conceptual implications and open problems}

While all the above characterizations consistently tie the absence of
GE to aspects of classicality as different as minimum state
complexity, classical simulatability, minimum uncertainty, and maximum
predictability, an independent notion of classicality is, in
principle, failure to violate appropriate {\em inequalities} violated
by quantum mechanics: for {\em pure} states, a Bell-type inequality is
violated if and only if the state is not separable in the conventional
sense \cite{Popescu92,Werner}.  Assessing whether the presence of GE
relative to appropriate observables may also be linked to the
violation of suitable ``generalized Bell inequalities'' represents,
from both a fundamental and philosophical perspective, one of the main
open questions about GE at present (see also \cite{Klyachko} for some
ideas along these lines).  From the point of view of QIS applications,
obtaining a more thorough resource-based characterization of GE,
linking the presence of GE to tasks which would not be achievable
otherwise, and/or quantifying the amount and type of GE required to
accomplish them, represent important areas for exploration where
additional fundamental insight about GE is expected to emerge.  In
particular, a resource-based approach may both further elucidate the
meaning of GE in a single quantum system \cite{Weinstein,Klyachko06}
as well as shed light on ways for distinguishing (conceptually and
operationally) between genuinely ``quantum'' and ``classical''
correlation aspects (see e.g. \cite{QCl,QCl1,QCl2}) within GE.

As we stressed throughout this work, the GE approach is naturally
suited to defining entanglement in settings more general than standard
quantum mechanics -- in particular, abstract operational theories
based on convex structures.  It may be worth observing that, by
abandoning objective, absolute notions of properties such as locality
and separability -- by acknowledging instead that different observers
can give different descriptions of these concepts and the ensuing
notion of entanglement, the GE approach also shares some of its
motivations with the recently proposed approach of {\em relational
quantum mechanics} \cite{Rovelli,Rovelli2}.  While important
differences between the latter and operational theories exist in terms
of how observers themselves are included and described, it could be
intriguing to further scrutinize differences and similarities between
the two approaches in the light of the GE concept.  Additional open
questions and implications for quantum foundations are discussed in
\cite{BarnumPRA03,BarnumFound}.

\section{Conclusion}

We believe that the richness of applications as well as the numerous
questions raised by GE program to date may speak by themselves about
the significance and potential of GE toward properly capturing the
unavoidable relativity of entanglement.  We hope that the GE
approach will provide fresh stimulus for the exploration of
entanglement to be extended into still-unexplored physical,
mathematical, and philosophical regions.

\section{Acknowledgments}

It is a pleasure to especially thank Manny Knill, Gerardo Ortiz, and
Rolando Somma for collaboration and interaction on various aspects of
the GE program.  L.V. is also indebted to Abner Shimony, Noah Linden,
and Sandu Popescu for thought-provoking discussions and exchange.
Work at Los Alamos is supported in part through the Laboratory
Directed Research and Development program.


\begin{thebibliography}{99}
\bibitem{note1} This scenario is realistic whenever the system can
only be accessed collectively in ensembles involving ``unconditional''
measurements on either a large ensemble of individual constituents or
a large number of repetitions starting from identical states, see also
\cite{BarnumPRA03}.
\bibitem{warn} We warn the reader that such states ($0$ aside) are
{\em not} extremal in the convex-sets sense, in the convex set $C$.
\bibitem{obs} Notice that operators commuting with $S_z
\leftrightarrow \hat{n}$ suffice for distinguishing states
$|\Phi^\pm\rangle$ from mixtures. The importance of taking into
account the constraint of fermion number conservation toward
consistently defining entanglement in states of identical fermions has
been independently stressed by Noah Linden (Private Communication).
\end{thebibliography}
\end{document}